\documentclass{sf2a-conf}
\usepackage{graphicx}

\begin{document}
\TitreGlobal{SF2A 2007}
\title{Chemodynamical evolution of interacting galaxies:\\ the GalMer view}


\author{P. Di Matteo} \address{LERMA, UMR 8112, CNRS, Observatoire de Paris, 61 Avenue de l'Observatoire, 75014 Paris,
France
}
\author{F. Combes$^1$} 
\author{I. Chilingarian$^1$}
\author{A. L. Melchior$^{1,}$}\address{Universit\'e Pierre et Marie Curie - Paris 6, 4 Place Jussieu, 75252 Paris Cedex 5, France} 
\author{B. Semelin$^{1,2}$}

\runningtitle{Chemodynamical evolution of interacting galaxies}

\setcounter{page}{1}

\index{Di Matteo P.}
\index{Combes F.} 
\index{Chilingarian I.}
\index{Melchior A. L.}
\index{Semelin B.}

\maketitle

\begin{abstract}
We have undertaken a large set of simulations of galaxy interactions and mergers (GalMer Project) in order to study the physical processes related to galaxy encounters. All morphological types along the Hubble sequence are considered in the initial conditions of the two colliding galaxies, with varying bulge-to-disk ratios and gas mass fractions. Different types of orbits are simulated, direct and retrograde, according to the initial relative energy and impact parameter. The self gravity of stars, gas and dark matter is taken into account through a tree-code algorithm, the gas hydrodynamics through SPH. Star formation is included adopting a density-dependent Schmidt law. This wide library of galaxy interactions and mergers, containing, at present, about 900 simulations of major encounters, represents an unique tool to investigate statistically the chemodynamical evolution of interacting systems. In the following, we present and discuss some results obtained exploring the dataset, together with some future perspectives.

\end{abstract}
%
\section{Introduction}


Galaxy interactions and mergers are responsible of reshaping galaxies from the Local Group to the limit of the observable Universe. They can produce  many spectacular phenomena: tidal tails and ripples, polar rings, kinematically decoupled cores,  star formation enhancements.  Understanding and interpreting all these phenomena is an ambitious task, complicated by the vaste range of physical parameters (morphology of the interacting galaxies, mass ratios, orbital parameters, etc,..) involved. In the last decade, numerical simulations have experienced important progresses, in terms of the adopted resolution and modelling of the relevant physical processes, such as gas physics, star formation and feedback (see, for example, the works by Mihos \& Hernquist (1994), Springel (2000) and Cox et al. (2006)).  While, on the one hand, these studies have been able to explain and reproduce many of the observable phenomena related to galaxy interactions, on the other hand, they have also shown that a large variety of results can be obtained. With the GalMer Project, we have chosen to adopt a statistical approach to the problem, by simulating and analysing thousands of simulations of interacting pairs. In this paper, we present some of the possible applications of this study, from the investigation of the physical mechanisms behind the large range of star formation enhancement found in interacting pairs, to the modelling of spectra of simulated galaxies, as well as to the kinematical properties of merger remnants. 

\section{The numerical model}
To model the galaxy evolution, we exploited a Tree-SPH code, in which gravitational forces are calculated using a hierarchical tree method (Barnes \& Hut 1986) and gas evolution is followed by means of smoothed particle hydrodynamics (Lucy 1977; Gingold \& Monaghan 1982). Each galaxy is made up of 120000 particles, distributed among gas, stars and dark matter, depending on the initial morphological type. For this first series of about 900 runs, we have 4 types of galaxies: E0, Sa, Sbc, and Sd, with a total mass of about $2\times 10^{11}M_{\odot}$,  12 different orbits, and two opposite senses on these orbits (i.e. direct and retrograde encounters). The two galaxy disks can have a relative inclination of $0^0$, $45^0$, $75^0$ or $90^0$. 
\section{A gallery of galaxy interactions}

\begin{figure}[h] 
 \begin{minipage}[b]{7.5cm}
\centering
\includegraphics[width=6.2cm,angle=270]{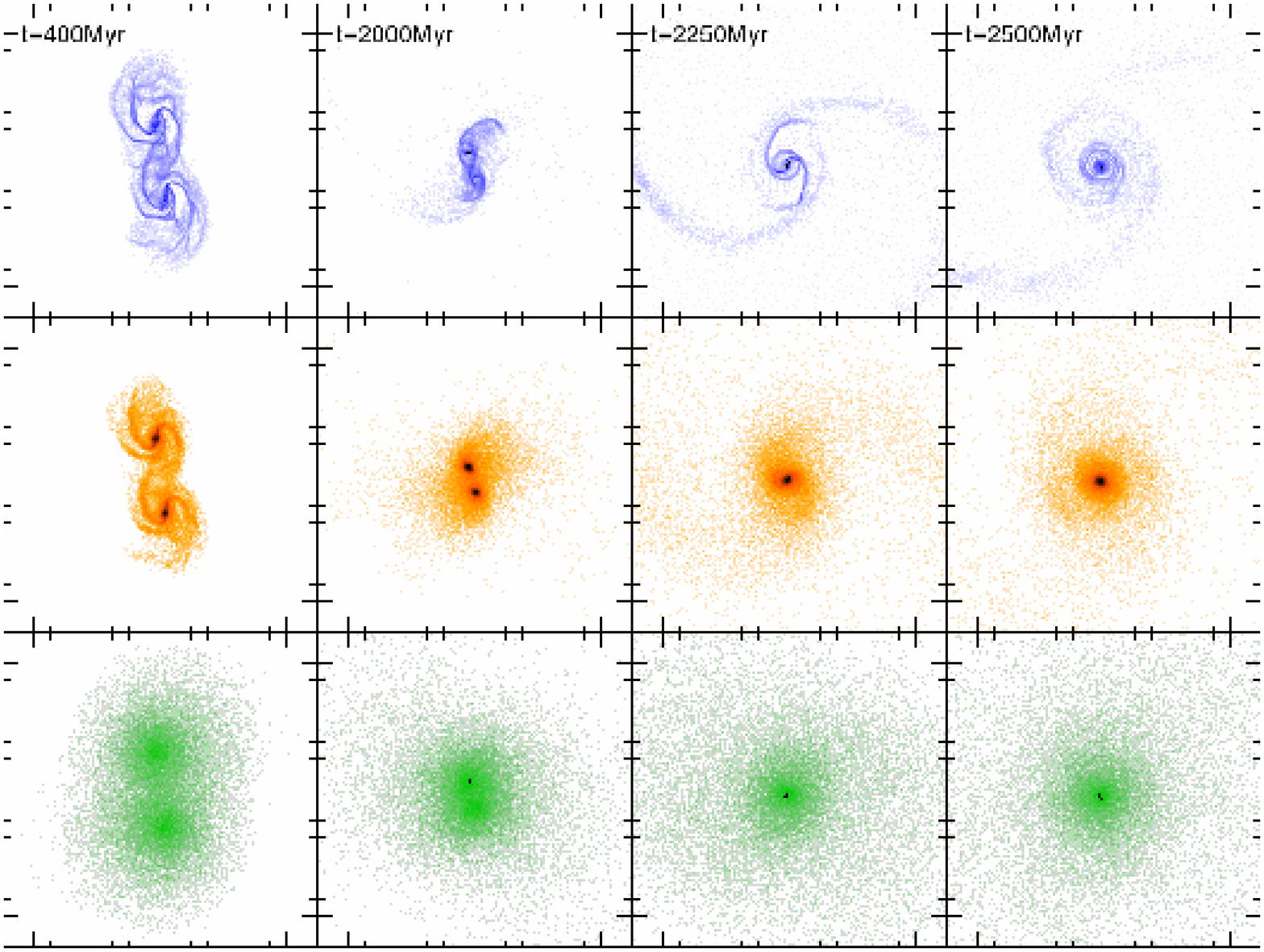} 
\end{minipage}
 \ \hspace{2mm} \hspace{3mm} \
 \begin{minipage}[b]{7.5cm}
\centering
\includegraphics[width=6.2cm,angle=270]{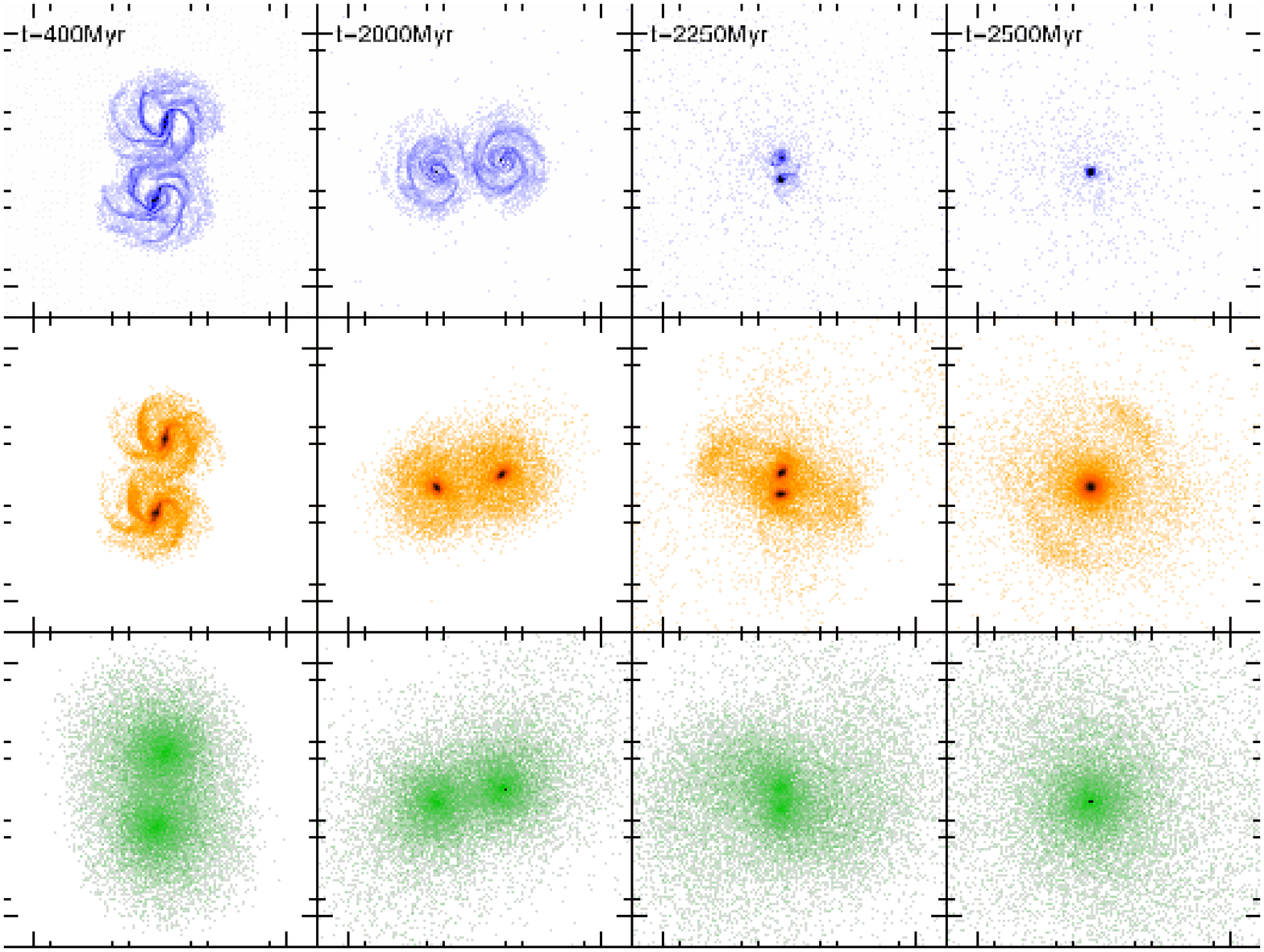} 
\end{minipage}
\caption{Evolution of the gas (upper panels), stars (intermediate panels) and dark matter (lower panels) during a direct (on the left) and a retrograde (on the right) merger between two gSb galaxies. Time is labelled in the upper part of the figure. Each frame is $50\times 50$ kpc in size. From Di Matteo et al (2007a).} 
\label{fig1} 
\end{figure} 

Fig.\ref{fig1} shows the encounters and successive mergers of two gSb galaxies, on direct (left panels) and retrograde (right panels) orbits, having coplanar disks. In both cases,  at the beginning of the simulation, the two galaxies are separated by a distance of 100 kpc. In the case of the direct encounter, as the two systems start to approach each other, they develop tails, populated by both stars and gas particles. The intense tidal field exterted during the first pericentre passage (t=400 Myr) leads also to a transfer of mass between the two systems. In turn, the retrograde interaction does not lead to any transfer of mass between the two galaxies, and also the formation of tidal tails is less obvious.   In many of the cases studied, the direct encounter leads a more rapid and dramatic merger compared to the retrograde one, with an expansion of the outer parts of the system. This can be explained by the fact that in retrograde encounters tides are less efficient, allowing most of the initial gas to stay in the spiral disk, rather than to be driven outwards. This great reservoir of gas can furnish the fuel for an intense burst of star formation in the merging phase. 
\section{Star formation in interacting pairs}

\begin{figure}[h] 
\centering
\includegraphics[width=6cm,angle=270]{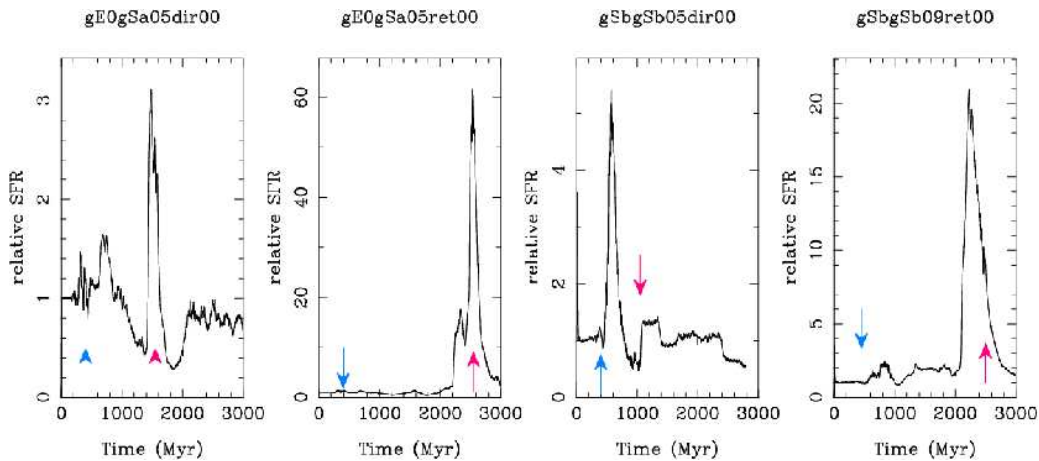} 
\caption{Star formation rate, versus time, for some galaxy mergers. The SFR is normalized to that of the corresponding isolated galaxies. The blue arrows indicate the first pericentre passage between the two galaxies, and the red arrows the merger epoch. See Di Matteo et al. (2007a).} 
\label{fig2} 
\end{figure}

Including star formation in gas dynamics is not a trivial task and a lot of different numerical methods and recipes can be adopted, in order to model, on the one hand, the star formation rate, and, on the other hand, to describe the effects that this star formation has on the surroundings. We adopted  the procedure described in Mihos \& Hernquist (1994), employing a ``local'' Schmidt law and an hybrid particles algorithm to implement it in our code. The resulting star formation rate evolution, for some galaxy encounters, is shown in Fig.\ref{fig2}.  Looking at these few examples, it appears obvious that interactions can increase star formation in the galaxy pairs from low levels (1.5-2 times the isolated case) to values typical of starburts galaxies (20-60 times the isolated case). In general, we found that retrograde encounters are more efficient than direct ones in driving star formation during the merger. This can be explained by the role played by tidal forces, particularly efficient for direct encounters in removing gas from the spiral disk (see Di Matteo et al. 2007a), reducing the available fuel for the burst phase. 

\section{Modelling spectra of interacting galaxies}
The numerical code adopted  allows also to trace the star formation history and the metal enrichment for every particle at any given snapshot of the simulation. Using the pre-computed grid of PEGASE.HR simple stellar populations  spectra and/or colours (Le Borgne et al. 2004), as well as the kinematics of the particles on the line of sight, we are able to make on-the-fly computation of spectra and/or colours for any region of the simulated galaxies as seen from any direction. The approach adopted to model these spectra is described in more detail in Chilingarian et al.,  this volume.
\section{Angular momentum redistribution and kinematical properties of merger remnants}

\begin{figure}[h] 
\centering
\includegraphics[width=5cm,angle=270]{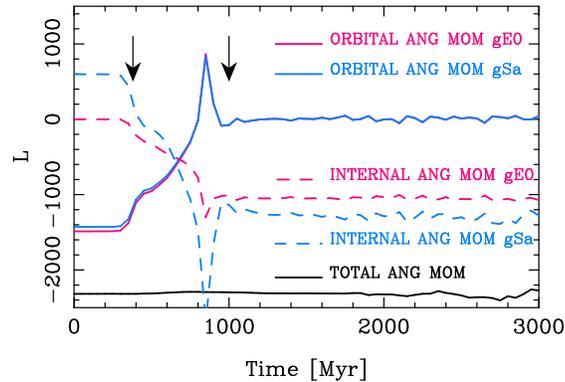} 
\caption{Evolution, with time, of the angular momentum $L$ during an elliptical-spiral Sa retrograde merger. Black line: total angular momentum; solid colored lines: orbital angular momentum of the elliptical (red curve) and of the Sa galaxy (blue curve); dashed colored lines: internal angular momentum of the elliptical (red curve) and of the Sa galaxy (blue curve). The angular momentum is in units of $2.3\times 10^{11} M_{\odot}\rm{ kpc kms^{-1}}$. The two arrows indicate the first pericenter passage and the merging time. From Di Matteo et al. (2007b).} 
\label{fig3} 
\end{figure}

\begin{figure}[h] 
\centering
\includegraphics[width=11cm,angle=0]{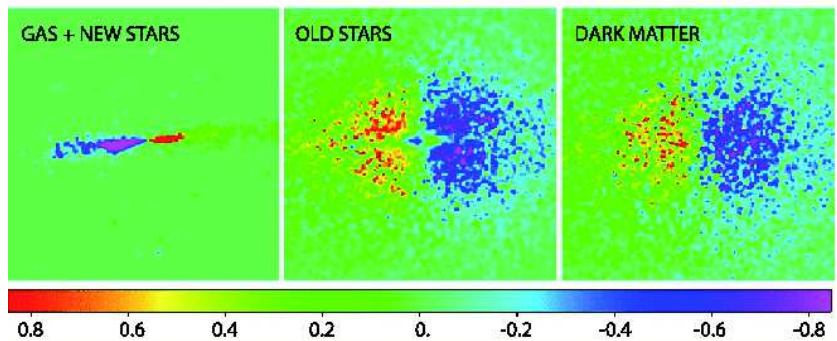} 
\caption{Line-of-sight velocity maps of gas (left panel), old stars (central panel) and dark matter (right panel) of the remnant of an elliptical-spiral merger. The maps are evaluated 1 Gyr after the coalescence of the two galaxies. Each side of the plot is 40 kpc in size.  Note the presence, in the old stellar population of the remnant galaxy, of a counter-rotating central core, whose extension is about 2 kpc in radius. Velocities are in units of 100 km/s. From Di Matteo et al. (2007b).} 
\label{fig4} 
\end{figure}

The tidal torques exterted on each galaxy during an interaction are also responsible of redistributing the initial orbital angular momentum between the different galactic components (i.e. gas, stars and dark matter). An example of the angular momentum transfer is given in Fig.\ref{fig3}, where the evolution with time of the total, orbital and internal angular momenta is given, for a retrograde encounter between an elliptical and a spiral Sa galaxy. This case is particularly instructive. It shows that the first transfer from orbital to internal angular momenta occurs at the first pericentre passage, when tidal torques remove part of the orbital angular momentum, converting it into internal spin of the two interacting systems.  This transfer continues until the two system approach the final stages of the merger event (t=1 Gyr): at that time the total angular momentum is completely distributed into the two systems, as internal rotation.\\
In this, as well as in other retrograde encounters, tidal torques are so efficient to be able to reverse the spin of the outer parts of the disk galaxy, while leaving its internal regions relatively unaffected. This ultimately determines the formation of  counter-rotating cores in the old stellar population of the merger remnant (see Fig.\ref{fig4}). The possibility to produce counterrotation from interactions between ellipticals and spiral galaxies has been only recently put in evidence (see Di Matteo et al. 2007b), and it represents a mechanism opposite to the one described in the literature up to the present (see, for example,  Balcells \& Gonzalez 1998).


\section{The GalMer database: providing the access to the scientific community}

The peculiarity of the GalMer Project does not reside only into the large number of detailed numerical simulations performed. 
Indeed, as part of the Horizon Project (see \emph{http://www.project-horizon.fr/}), it will provide to the scientific community a free and user-friendly access to all the runs performed, together with the possibility to make on-the-fly analysis of the runs. In particular, it will be possible:
to produce maps of gas, stars and dark matter, at different stages of the interactions, as seen from different directions;
  to trace the star formation histories and metal enrichment during galaxy encounters, and to compare these evolutions to those of the corresponding galaxies evolved in isolation; 
to compute spectra and colors for any region of the simulated galaxies, as seen from any direction.



\end{document}